\documentclass{article}

\usepackage[a-1b]{pdfx}
\usepackage{graphicx}
\usepackage{amsmath}
\usepackage{subfigure}
\usepackage{chemarrow}
\usepackage{draftwatermark}

\newcommand{\attrib}[1]{%
\nopagebreak{\raggedleft\footnotesize #1\par}}

\graphicspath{./}

\newcommand{\pr}{p_{0}}
\newcommand{\pt}{p(t_m)}

\newcommand{\prn}{p_{0,i+1}}
\newcommand{\pc}{p_{cr}}
\newcommand{\pred}{p_{r}}
\newcommand{\predt}{p_{r}(t_M)}
\newcommand{\predi}{p_{r,i}}
\newcommand{\predn}{p_{r,i+1}}

\newcommand{\Vr}{V_{m,0}}
\newcommand{\Vt}{V_m(t_m)}

\newcommand{\Vc}{V_{m,cr}}
\newcommand{\Vred}{V_{m,r}}
\newcommand{\Vredt}{V_{m,r}(t_M)}
\newcommand{\Vredi}{V_{m,r,i}}
\newcommand{\Vredn}{V_{m,r,i+1}}

\newcommand{\T}{T}
\newcommand{\Tr}{T_{0}}
\newcommand{\Tt}{T(t_m)}

\newcommand{\Tc}{T_{cr}}
\newcommand{\Tred}{T_{r}}
\newcommand{\Tredt}{T_{r}(t_M)}
\newcommand{\Tredi}{T_{r,i}}
\newcommand{\Tredn}{T_{r,i+1}}

\newcommand{\dFdx}{\frac{\delta F}{\delta x}}
\newcommand{\dFdy}{\frac{\delta F}{\delta y}}

\newcommand{\dxdt}{\frac{d x}{d t_M}}
\newcommand{\dydt}{\frac{d y}{d t_M}}
\newcommand{\dxdy}{\frac{d x(t_M)}{d y(t_M)}}
\newcommand{\quotyx}{\frac{\dFdy}{\dFdx}}

\newcommand{\dFdplong}{(3 \Vred^3 - \Vred^2)}
\newcommand{\dFdVmlong}{(9 \pred \Vred^2 - 2 (\pred + 8 \Tred) \Vred + 9)}
\newcommand{\dFdTlong}{(- 8 \Vred^2)}

\newcommand{\vvi}{\vec{v}_i}

\begin{document}
\SetWatermarkText{TD-VDW-PF  \today}

\title{Simulating phase transitions by means of quasi static state changes: the
capabilities of the time dependent Van der Waals equation of state}

\author{Peter Friedel\\ \\
	Leibniz-Institut f{\"u}r Polymerforschung Dresden e.V.\\
        Hohe Str. 6\\
        D-01069 Dresden\\
        Tel.: +49-351-4658751\\
        Fax: +49-351-4658752\\
        email: friedel@ipfdd.de
}

\maketitle

\abstract{
The Van der Waals equation (VdW-EoS) is a prototype equation of state for
realistic systems, because it contains the excluded volume and the particle
interactions. Additionally, the simulated annealing (and the similar simulated
compressing) approach applies the time dependence on to one of the variables
of state to simulate quasi static state changes. The combination of both
enables the simulation of time dependent processes like phase transitions of
subcritical, critical and supercritical substances on every arbitrary 
condition including a passage over points of singularity of the corresponding
susceptibility coefficients. This is achieved by a new simulation approach
called simulated expansion. This approach makes the simulation comparable to
natural processes which exhibit gradual changes in volume, rather than changes
in temperature or pressure, as exercised in simulated annealing or
compressing. The demonstrated method here serves as a blue print for more
general classes of simulation approaches.
}

\newpage

\tableofcontents

\section{Introduction}
\label{intro}

\begin{quote}
\emph{What is this subject in its essence and in its properties here? What is
it a substance for? What are the kind of forces acting in it? What is it doing
within this world and how long will be its duration?}
\end{quote}
\attrib{Marc Aurel (121 - 180, C.E.)~\cite{MA-selbstbetrachtungen-2011}}

Equations of state are playing an important role within a lot of simulation
methods because they characterize the properties of a system in a unique way
(see for example St{\"o}cker~\cite{ma:stoecker-2004} or
Haberlandt~\cite{md:molekulardynamik-1994}).

One of these analytic equations of state is the Van der Waals equation.
Although, it was presented and published more than a hundred years
ago~\cite{JV-model-phdthesis-1873}, it still represents the prototype of
an equation of state because it introduces the excluded volume and the
interactions between the particles of the system for the first time. 
These two important issues are still included in many other simulation 
techniques, because such a model is very useful for describing a wide area of 
real compound systems. Therefore, Van der Waals did provide an 
important contribution to the understanding of the
behavior of realistic systems by the help of his equation
(see also the list of Nobel prize winners in physics: J.D. v.d. Waals,
1910~\cite{AK-nobelpreise-physik-1900-1910}). Despite, or perhaps because of 
its simplicity, this equation is applied frequently until today
(for example see Parneix~\cite{ev:parneix-1998},
Berberon-Santos~\cite{MB-vanderwaals-equation-solutions-article-2008} and
Sta\v{s}kiewicz~\cite{BS-vanderwaals-equation-article-2014}).

The corresponding approaches of simulation, either analytically or 
numerically, which are making use of the Van der Waals ansatz of excluded 
volume and particle interaction are wide spread and may be distinguished from each
other by the time scale or by the method in which the energy is calculated.

Performing simulations of such realistic systems independent on the system
size, the time scale or the method of calculating the energy, a general and 
common procedure may be denoted which works as follows:

\begin{enumerate}
\item The first step is the system setup, either abstract or specifically.
\item In a second step the system is relaxed to a local energetic
minimum, which could be performed by gradient or variational calculation
methods~\cite{ma:bronstein-2013}. The resulting state of minimum energy is a
well defined one and depends on the configuration of the starting system.
\item The application of methods which allow a heat or pressure bath coupling
(see, e.g., Andersen~\cite{md:andersen-1980},
Berendsen~\cite{md:berendsen-1984}, 
Nose-Hoover~\cite{SN-nose-hoover-1984,WH-nose-hoover-1985} and others) 
enables a relaxation of the system to states of given macroscopic reference
temperatures or pressures.
\item After this procedure, the equilibrated system may be simulated further 
for data production, until a sufficiently large ensemble is obtained.
\item Alternately, the state of such an equilibrated system may be modified
by means of so called quasi static state changes, i.e., the
''Simulated Annealing'' - SA - (see Kirkpatrick~\cite{md:kirkpatrick-1983})
or the so called ''Simulated Compressing'' - SC - method (see e.g.
Zhao~\cite{JZ-simulated-compression-article-2013}).
\end{enumerate}

Nature provides a lot of important processes which may be
studied by experimental and/or by simulation approaches followed by a
comparison of the results of both strategies. The most important
of such processes are phase transitions of first or second order, i.e.,
melting/crystallization,
sub\-li\-ma\-tion\-/\-re\-sub\-li\-ma\-tion or evaporations/condensations, but also
solid phase transitions or order/disorder transitions in polymer blends or
block co-polymer systems (see also Landau~\cite{th:landau5-1987},
therein the chapter XIV), and among others. The theoretical understanding of these
processes is still a big challenge.

This challenge can be illustrated by an example: Along with the simulation of
such natural processes like phase transitions with the purpose of comparing them
with the corresponding experimental discoveries some unexpected difficulties
may appear. E.g., different computer experiments of NpT ensembles tried to
simulate the process of the phase transition from liquid to gaseous water
(see Walser~\cite{RW-md-simulation-water-article-2000},
Dou~\cite{md:dou-2001,md:dou-2-2001} and Zahn~\cite{md:zahn-2004})
by means of a linear increase of the temperature. These simulations had to be
stopped always close to the spontaneous phase transition point near 500 K at
101.3 kPa of pressure. Vice versa, corresponding simulations of an NVT
ensemble (V - constant volume instead of constant pressure) could be
performed successfully (see Neimark~\cite{ev:neimark-2005} or
Medeiros~\cite{MM-GEMC-water-article-1997}).

The reason why the simulation of the NpT ensemble of the water evaporation 
process was failing seems to lie within the expansion coefficient. This
coefficient reaches obviously singularities below or close to the critical
point when passing the phase transition point applying isobar conditions. 
Therefore, such theoretical investigations are impracticable, and successful
simulations in that sense are not known. On the other side, the natural process
performs such a transition without any hurdles. Similar observations could be
made in melting or crystallization processes on isotherm conditions
(see Albano~\cite{CA-isotherm-crystallization-kinetics-article-2003} and some
corresponding NVT molecular dynamics simulations of
Baidakov~\cite{VB-md-simulation-isotherm-crystallization-article-2010}).
In this case, the compressibility as the responsible susceptibility 
coefficient shows a similar behavior like the expansion coefficient.

In summary, it has to be stated that there is a gap between the natural 
observations and the corresponding theoretical investigations according 
this subject. So, it is the not quite easy challenge and aim of this 
manuscript to find out an opportunity of closing this vacancy. Otherwise, 
it is the trial to get an overview about all the opportunities to perform
quasi static state changes on the overall palette of thermodynamic conditions, 
independent from the degree of abstraction of the substances, the applied 
time or energy scale or the type of collected statistical ensembles.

The manuscript should be organized as follows.
At first, the Van der Waals equation is described in it's reduced
implicit form. A reasoning is given for the introduction of the
time dependence to the VdW-EoS at the macroscopic time scale. Then
this time dependence is applied and the total time differential
is derived. Afterwards, some special mathematical considerations
are described which are important for the later discussion of three
special cases. This discussion includes some considerations about
the coherence between the Van der Waals equation and the Maxwell
construction~\cite{JM-maxwell-construction-1875}. The conclusions
summarizes the systematic collection of new options of quasi static state
changes which are now available to simulate such interesting challenges like
phase transitions on any arbitrary conditions, i.e., including the famous
circular Carnot process (see in Stingers~\cite{IS-posthumanities-inbook-2010}).

\section{Quasi static state changes}
\label{sec:vdw}

\subsection{The microscopic thermodynamic quantities of
state $\Tt$, $\pt$ and $\Vt$ vs. the macroscopic reference values $\Tr$, 
$\pr$ and $\Vr$ in molecular dynamics simulations}
\label{subsec:vdw-thermodynamics}

Molecular dynamics simulations demonstrate the time dependence of the 
microscopic thermodynamics quantities of state. A general and useful overview
is given by van Gunsteren~\cite{md:gunsteren-1990} or 
Haberlandt~\cite{md:molekulardynamik-1994}. Additionally, 
Berendsen~\cite{md:berendsen-1984} developed his method of heat and pressure 
bath coupling  to relax the molecular dynamics ensemble to the given
macroscopic reference temperature $\Tr$ and pressure $\pr$. Nevertheless,
these macroscopic reference values may be changed slowly during time. This is
called a quasi static state change (e.g., see 
Baehr~\cite{HD-quasistatischer-zustand-article-1961} and
Guggenheim~\cite{EG-principle-corresponding-states-article-1945}) like
the linear SA approach (see Kirkpatrick~\cite{md:kirkpatrick-1983}).

The scaling of the velocities $\vvi (t_m)$ at the time scale $t_m$ (the lower
$m$ means the microscopic time scale) by the scaling factor $\lambda$
(see e.g. Berendsen~\cite{md:berendsen-1984}, Formula (34))

\begin{equation}
\label{equ:berendsen34}
\lambda (t_m) = \Bigg[ 1 + \frac{\Delta t_m}{\tau_T} 
\Bigg\{ \frac{\Tr}{\T {(t_m - 0.5 \Delta t_m)} } - 1 \Bigg\} \Bigg]^{\frac{1}{2}} 
\end{equation}

enables the relaxation of the system to a given absolute macroscopic reference
temperature $\Tr$, changeable slowly by the linear SA approach, and the system
is following that reference temperature.

By the help of a value called $\mu$ (or more general a corresponding tensor)

\begin{equation}
\label{equ:my}
\mu (t_m) = 1 - \frac{\Delta t_m}{\tau} (\pr - \pt)
\end{equation}

with $\tau$ as a pressure coupling constant, $\pr$ as a macroscopic reference
pressure (in scalar form here, or in a general tensorial form otherwise, 
slowly changeable in time by the SC 
method~\cite{JZ-simulated-compression-article-2013}) the ensemble volume of
the next time step is given by

\begin{equation}
\label{equ:volume-next}
V(t + \Delta t_m) = \mu (t_m) V(t_m).
\end{equation}

Please note, that the volume depends on the microscopic pressure $\pt$,
the macroscopic reference pressure $\pr$, the microscopic temperature $\Tt$
and the macroscopic reference temperature $\Tr$. These four values are 
independent variables of state in such an equilibrated ensemble controlled by
the heat and pressure bath coupling method of Berendsen. Therefore, the 
microscopic volume $\Vt$ of the simulation box is {\bf not} an independent variable
of state under these circumstances. $\Vt$ fluctuates around a macroscopic
reference volume $\Vr$ what is the average result of the equilibration process.

\subsection{The time dependent Van der Waals equation in its implicit reduced
form}
\label{subsec:vdw-reduced}

The reduced implicit form of the Van der Waals equation results by the
application of the principle of the corresponding states
(see Guggenheim~\cite{EG-principle-corresponding-states-article-1945}) to:

\begin{equation}
\label{equ:vdw-reduced}
F_{VdW,r}(\pred, \Vred, \Tred) = ({\pred + \frac{3}{\Vred^2}}) 
({3 \cdot \Vred - 1}) -  8 \cdot \Tred = 0
\end{equation}

Of course, it has to be taken into account the consideration about the validity
of the Van der Waals equation, that the $\Vred$ has to be always greater than
$\frac{1}{3}$.

This equation of state describes the behavior of a molecular ensemble with
a more or less large number $N$ of molecules. If the number $N$ goes to $\infty$
then this is called the thermodynamic limit, and the thermal fluctuations of 
the global quantities are negligible. In this
microscopic sense, the equation of state is no longer time dependent.

What should be shown and discussed here is the introduction of a time
dependence into an equation of state in a macroscopic sense, i.e.,
at a quite different time scale $t_M$ (upper $M$ stands for macroscopic
time scale) than the molecular system resides (time scale $t_m$).

It is the time scale $t_M$ of some every day observations, e.g., the cooking of
the coffee water in the morning, which should be considered here. The 
temperature $\Tr$ of Eq. (\ref{equ:berendsen34}) and the pressure $\pr$ of
Eq. (\ref{equ:my}) represent such macroscopic quantities which are changing
slowly at the time scale $t_M$ in the sense of quasi static state changes. 
The microscopic ensemble at the microscopic time scale $t_m$ itself is still
residing at an equilibrated state according Eq. (\ref{equ:vdw-reduced}).

This important aspect of the given considerations makes the introduction
of time $t_M$ into an equation of state not only to be justified but also almost to
be required. The consequences which are following and shown in the next sections
are as significant as they are simple in their mathematics.

To perform quasi static state changes Eq. (\ref{equ:vdw-reduced}) can be
written in it's time dependent form:

\begin{eqnarray}
\label{equ:vdw-reduced-expanded}
& & F(\predt, \Vredt, \Tredt) \nonumber \\
& = & \Vredt^2 \cdot F_{VdW,r}(\predt, \Vredt, \Tredt) = 0
\end{eqnarray}

It is important to describe the physical meaning of that equation. At first,
the implicit description still shows the character to be an equation of state,
i.e., it describes an equilibrated state.
Secondly, the time dependence shows the dynamics of the macroscopic
systems behavior in the sense of a quasi static state variation. 
The described dynamic is the dynamic of at least two different opposite
processes which are going from one state into another and vice versa (e.g., 
the evaporation/condensation or melting/crystallization etc.) building
a dynamic equilibrium in the whole.

The specific total time differential of
Eq. (\ref{equ:vdw-reduced-expanded}) looks like:

\begin{eqnarray}
\label{equ:vdw-implicit-total-differential-time-dependent}
0 & = & \dFdplong \frac{d \predt}{d t_M} \nonumber \\
  & + & \dFdVmlong \frac{d \Vredt}{d t_M} \nonumber \\
  & + & \dFdTlong \frac{d \Tredt}{d t_M}
\end{eqnarray}

and, of course, the three ordinary time
differentials of the thermodynamic quantities $\Tredt$, $\predt$
and $\Vredt$ as the macroscopic reference values in reduced form
(shortly written as $\Tred$, $\pred$ and $\Vred$ in the
following from here) can be derived immediately:

\begin{eqnarray}
\label{equ:dqdt-T}
\frac{d \Tred}{d t_M} & = & - \frac{\frac{\delta F}{\delta \Vred} \frac{d \Vred}{d t_M} + 
\frac{\delta F}{\delta \pred} \frac{d \pred}{d t_M}}{\frac{\delta F}{\delta \Tred}} \\
 & = & - \frac{\dFdVmlong}{\dFdTlong} \frac{d \Vred}{d t_M} \nonumber \\
 &   & - \frac{\dFdplong}{\dFdTlong} \frac{d \pred}{d t_M} \nonumber
\end{eqnarray}

\begin{eqnarray}
\label{equ:dqdt-p}
\frac{d \pred}{d t_M} & = & - \frac{\frac{\delta F}{\delta \Vred} \frac{d \Vred}{d t_M} + 
\frac{\delta F}{\delta \Tred} \frac{d \Tred}{d t_M}}{\frac{\delta F}{\delta \pred}} \\
 & = & - \frac{\dFdVmlong}{\dFdplong} \frac{d \Vred}{d t_M}\nonumber \\
 & &   - \frac{\dFdTlong}{\dFdplong} \frac{d \Tred}{d t_M}\nonumber 
\end{eqnarray}

\begin{eqnarray}
\label{equ:dqdt-V}
\frac{d \Vred}{d t_M} & = & - \frac{\frac{\delta F}{\delta \pred} \frac{d \pred}{d t_M} + 
\frac{\delta F}{\delta \Tred} \frac{d \Tred}{d t_M}}{\frac{\delta F}{\delta \Vred}} \\
 & = & - \frac{\dFdplong}{\dFdVmlong} \frac{d \pred}{d t_M} \nonumber \\
 &   & - \frac{\dFdTlong}{\dFdVmlong} \frac{d \Tred}{d t_M} \nonumber
\end{eqnarray}

with

\begin{equation}
\label{equ:reducing}
\Tred =\frac{\Tr}{\Tc} \quad \pred = \frac{\pr}{\pc} \quad \Vred =\frac{\Vr}{\Vc}
\end{equation}

, $\Tc$, $\pc$ and $\Vc$ as the material specific values of the critical
point including $\Tr$ and $\pr$ as the macroscopic reference values
according Eq.s (\ref{equ:berendsen34}) and (\ref{equ:my}) including $\Vr$ as
the resulting equilibrium volume.

\section{Mathematical considerations}
\label{sec:math}

\subsection{Curve sketching of the time differentials}
\label{subsec:curve-sketching}

\subsubsection{Three special cases}
\label{subsubsec:special-cases}

Before discussing the set of Eq.s (\ref{equ:dqdt-T}), (\ref{equ:dqdt-p}) and
(\ref{equ:dqdt-V}), it is necessary to reduce the complexity by
setting one of the time differentials called $\frac{d z}{d t_M}$ to $0$. 
Afterwards, at first three special cases and at second the general cases can be
discussed.

The setting $\frac{d z}{d t_M}$ to $0$ simplifies Eq.s (\ref{equ:dqdt-T}), 
(\ref{equ:dqdt-p}) and (\ref{equ:dqdt-V}) to the general form:

\begin{equation}
\label{equ:diff}
\dxdt = - \Bigg({\quotyx}\Bigg)_{z} \dydt
\end{equation}

for every of the three main cases at all:

\begin{eqnarray}
\label{equ:threecases}
	x (y, z = const.) & = & \pred(\Tred, \Vred = const ), 
\nonumber \\
	& & \Vred(\Tred, \pred = const ), 
\nonumber \\
	& & \Vred(\pred, \Tred = const ) 
\end{eqnarray}

It is quite clear that the expected kind of time dependence of these functions 
is defined by the fracture of the corresponding partial differential quotients
in Eq. (\ref{equ:diff}) (see Eq. (\ref{equ:vdw-implicit-total-differential-time-dependent})),
i.e., ambiguous, unambiguous and monotonic function types are possible.

The ambiguous function type is, for example, not integrable what is a big hindrance for
the further characterization and usability. One opportunity of handling such a function
is the detour along the inverse function. In that sense, the hint No. 2 within the mathematical
textbook of Bronstein~\cite{ma:bronstein-2013} at page 52 (Cite: ''If there is a
function with non-monotonic behavior separable into monotonic pieces then the
corresponding inverse function exists for every of that monotonic pieces.'')
becomes importance: it is possible to transfer every given function
into it's inverse one at the full domain and range areas and vice versa.
In other words: If a function is ambiguous and therefore, not integrable, the
corresponding inverse function is the wanted detour.

\subsubsection{Functions and their inverse functions}
\label{subsubsec:math-inverse}

The well known relation between the derivative of a function and the
derivative of the corresponding inverse function

\begin{equation}
\label{equ:derivative-inverse}
\dxdy = \frac{1}{\frac{d y(t_M)}{d x(t_M)}}
\end{equation}

is helpful in the subject under investigation.

From the first subset (the three main cases of Eq. (\ref{equ:threecases})) 
the exchange of $x$ by $y$ delivers promptly a second subset, i.e., the
corresponding inverse functions:

\begin{eqnarray}
\label{equ:pT}
\pred(\Tred, \Vred = const) \quad and \quad \Tred(\pred, \Vred = const)\\
\label{equ:TV}
\Vred(\Tred, \pred = const) \quad and \quad \Tred(\Vred, \pred = const)\\
\label{equ:Vp}
\Vred(\pred, \Tred = const) \quad and \quad \pred(\Vred, \Tred = const)
\end{eqnarray}
.

The curve sketching of this set of formulas (\ref{equ:pT}),
(\ref{equ:TV}) and (\ref{equ:Vp}) shows all function types for both subsets:

\begin{enumerate}
\item monotonic increasing: $\pred(\Tred, \Vred = const)$ and $\Tred(\pred, \Vred = const)$;
\item unambiguous with extrema: $\Tred(\Vred, \pred = const)$ and
$\pred(\Vred, \Tred = const)$;
\item ambiguous: $\Vred(\Tred, \pred = const)$ and
$\Vred(\pred, \Tred = const)$;
\end{enumerate}

An ambiguous function can be exchanged by the corresponding
unambiguous function with extrema, which is integrable now.

\subsubsection{Integrability}
\label{subsubsec:integration}

In the case that the fractures in Eq. (\ref{equ:diff}) have an unambiguous or monotonic
character and no poles exist it is possible to continuous with integration.

The integration may be performed in two different ways. The first one is quite
simple by removing the time to find an analytic solution with the aim to discuss
the corresponding curvature.

Eq. (\ref{equ:diff}) can be written in integral form and analytically solved:
\begin{equation}
x_{i+1} = x_i - \int_{y_i}^{y_{i+1}} \Bigg({\quotyx}\Bigg)_{z} d y
\end{equation}

The numerical integration opens up the opportunity to hold the change of
$y$ constant by the definition of $k_{dy}$:

\begin{equation}
\label{equ:consty}
\frac{d y}{d t_M} = const. = k_{dy}
\end{equation}

what gives immediately

\begin{equation}
\label{equ:ykdyt}
y_{i+1} = y_i + k_{dy}  (t_{i+1,M} - t_{i.M})
\end{equation}

and then Eq. (\ref{equ:diff}) looks like

\begin{equation}
\label{equ:diffkdy}
\dxdt = - \Bigg({\quotyx}\Bigg)_{z} k_{dy}
\end{equation}

Setting the differential quotient $\dxdt$ equal to a forward difference
quotient~\cite{ma:schwarz-2004}, the numerical integration
with $\Delta t_M = (t_{i+1,M} - t_{i,M})$ follows:

\begin{equation}
\label{equ:xydifference}
\dxdt = \frac{x_{i+1} - x_i}{t_{i+1,M} - t_{i,M}} + O (\Delta t_M)
\end{equation}

with the numerical error term $O (\Delta t_M)$.

This is the Euler polygon approach which should be used for simplicity, here,
\begin{equation}
\label{equ:numericxt}
x_{i+1} \approx x_i - \Bigg({\quotyx}\Bigg)_{z,t_{i,M}} k_{dy} \Delta t_M 
\end{equation}

with order of error 1. This numerical integration method, applying the time 
step $\Delta t_M = 1.0$, serves as an offer for the usage in Eq.s 
(\ref{equ:berendsen34}) and (\ref{equ:my}) for providing $\Tr$ and $\pr$ values.

\subsection{The susceptibility coefficients}

A second simplification of Eq. (\ref{equ:diff}) delivers immediately

\begin{eqnarray}
\label{equ:diffot}
\phi (x(y(t_M))) & = & \frac{1}{x(y(t_M))} \dxdy \nonumber \\
& = & \frac{d \; ln x(y(t_M))}{d y(t_M)} \nonumber \\
& = & - \frac{1}{x(y(t_M))} \Bigg({\quotyx}\Bigg)_{z}
\end{eqnarray}
.

This equation describes directly the definition of all three important
susceptibility coefficients of the tension, the expansion and the compressibility,
summarized as $\phi$ in general. According the subsection above it is allowed to
define also susceptibility coefficients of the corresponding inverse functions.
For shortening, the braces of direct or indirect time dependencies
are no longer written.

\section{Results and Discussion}
\label{sec:disc}

According the simplification given in Eq. (\ref{equ:diff}), three different 
conditions can be distinguished:

\begin{equation}
\label{equ:cond}
\frac{d \Vred}{d t_M} = 0 \qquad \frac{d \pred}{d t_M} = 0 
\qquad \frac{d \Tred}{d t_M} = 0
\end{equation}

, i.e., the isochor, the isobar or the isotherm case. Every of that cases
delivers two different subsets. With the exception of the isochor case the
first subset always describes the well known situation of an ambiguous function
including the consequences to the corresponding susceptibility coefficient.
This is the situation where the appropriate trials of simulations always
had to be stopped close to the evaporation point of the subcritical
liquid because the volume is running into unlimited fluctuations. This is the
reason why the first subset does not allow to pass the point of
phase transition.

Therefore, the here applied calculation procedure is the following:

\begin{enumerate}
\item Starting with the calculations of the dependent variable of state of 
the {\bf second} subset according Eq. (\ref{equ:numericxt});
\item Calculating the susceptibility coefficients of the {\bf second} subset
according Eq. (\ref{equ:diffot});
\item Calculating the susceptibility coefficients of the {\bf first} subset by the
help of the relation Eq. (\ref{equ:derivative-inverse});
\item Using the inverse presentation of the results of Eq. (\ref{equ:numericxt})
for graphical representation of the dependent variable of state of the {\bf first}
subset;
\end{enumerate}

Every of the following three subsections includes one figure containing
four sub-figures. According the calculation procedure, the sub-figures are
presented in two columns and two rows. The left column shows the first subset
(sub-figures (a) and (c)), the right one the second subset (sub-figures
(b) and (d)). The first row presents the corresponding
dependent variable of state from the independent one (sub-figures (a) and (b)), 
and the second row represents the corresponding susceptibility coefficient
(sub-figures (c) and (d)).

In this way the seemingly unsolvable problem of the first subset, 
discussed at first in the following three subsections, can be solved in every
case using the detour along the inverse function, i.e., along the second subset.

\subsection{The isochor case}
\label{subsec:isochor}

The set of Eq.s (\ref{equ:dqdt-T}), (\ref{equ:dqdt-p}) and (\ref{equ:dqdt-V})
including the condition $\frac{d \Vred}{d t_M} = 0$ leads to:

\begin{eqnarray}
\label{equ:dqdt-isochor-1}
\frac{d \pred}{d t_M} & = & - \Bigg(\frac{\frac{\delta F}{\delta \Tred}}
{\frac{\delta F} {\delta \pred}}\Bigg)_{\Vred} \frac{d \Tred}{d t_M}\\
 & = & \frac{8}{(3 \Vred - 1)} \frac{d \Tred}{d t_M} \nonumber \\
\label{equ:dqdt-isochor-2}
\frac{d \Tred}{d t_M} & = & - \Bigg(\frac{\frac{\delta F}{\delta \pred}}
{\frac{\delta F}{\delta \Tred}}\Bigg)_{\Vred} \frac{d \pred}{d t_M}\\
 & = & \frac{(3 \Vred - 1)}{8} \frac{d \pred}{d t_M}\nonumber
\end{eqnarray}

The constant reduced molar volume $\Vred$ as a parameter can be chosen either 
as a subcritical (e.g. $\Vred = 0.75$), a critical (i.e., $\Vred = 1.00$)
or a supercritical value (e.g. $\Vred = 1.33$).

\begin{figure*}[ht]
\begin{center}
\subfigure[$\pred$ over $\Tred$ (acc. Eq. (\ref{equ:vdw-reduced}))]
{\resizebox{0.39\textwidth}{!}
{\includegraphics{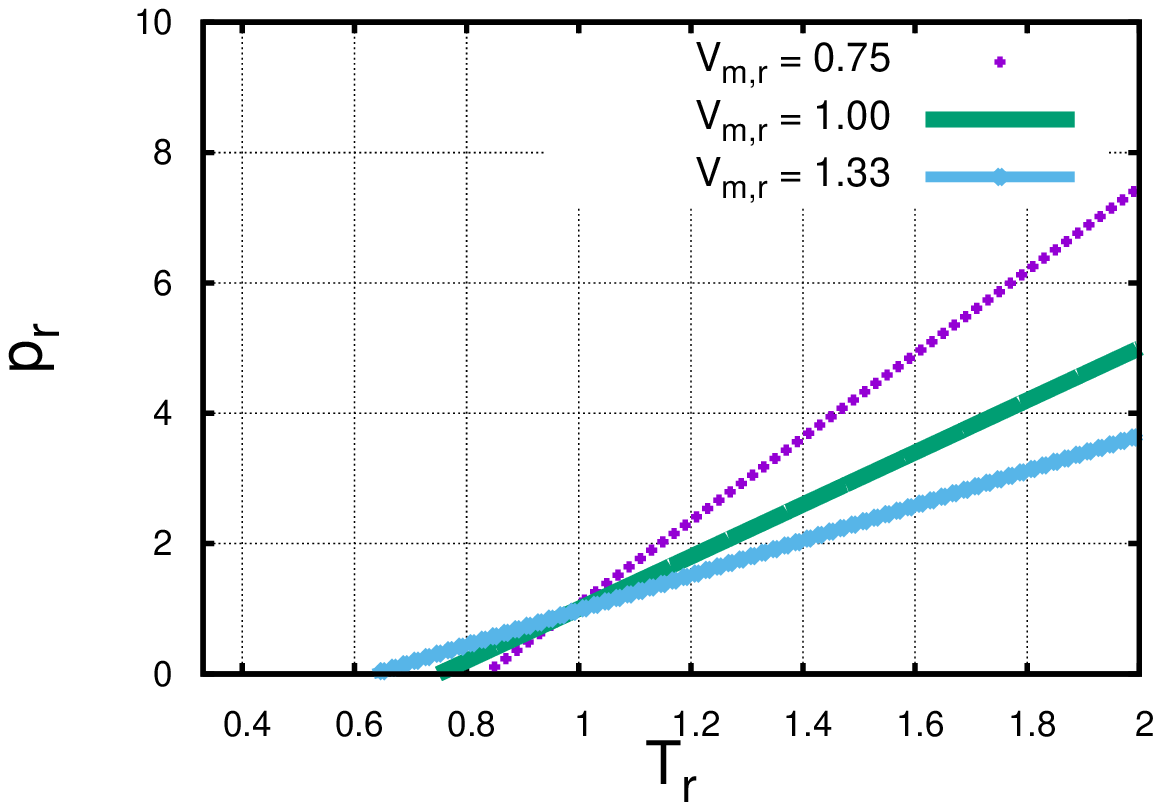}}}
\subfigure[$\Tred$ over $\pred$ (acc. Eq. (\ref{equ:vdw-reduced}))]
{\resizebox{0.39\textwidth}{!}
{\includegraphics{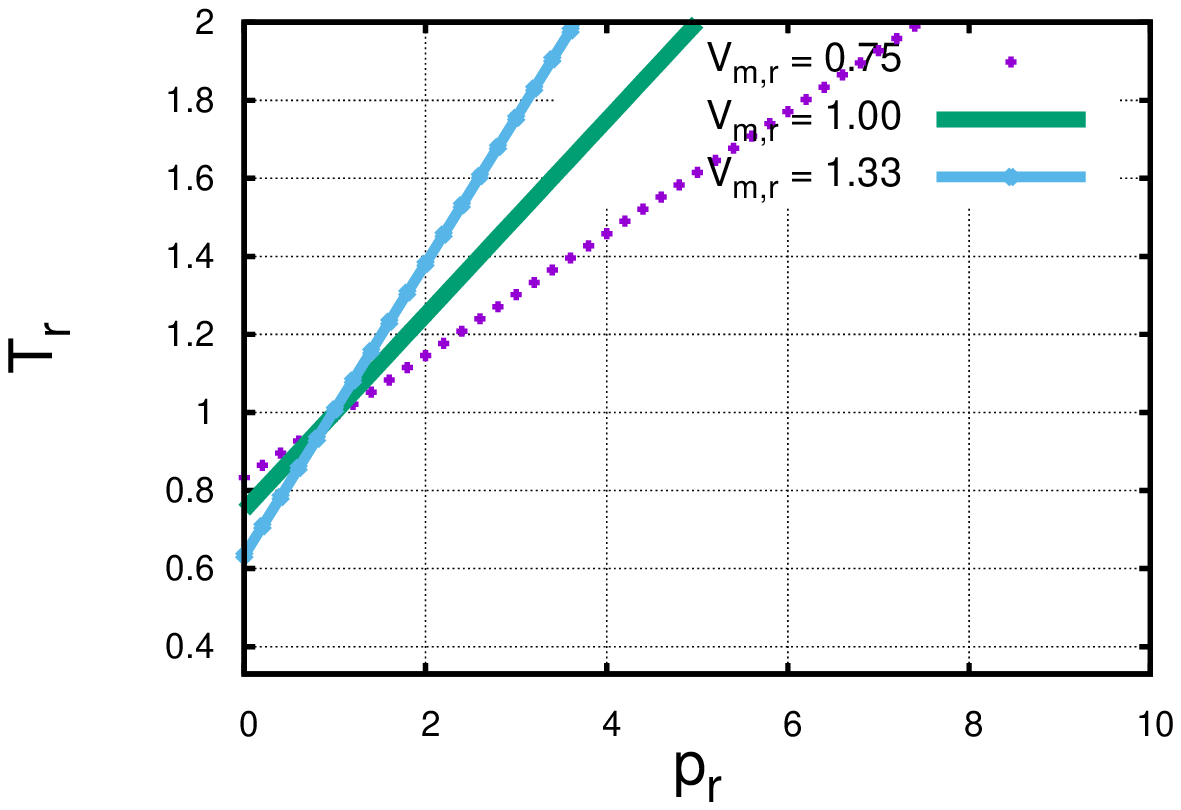}}}
\subfigure[$\beta_r$ over $\Tred$ (acc. Eq. (\ref{equ:tension}))]
{\resizebox{0.39\textwidth}{!}
{\includegraphics{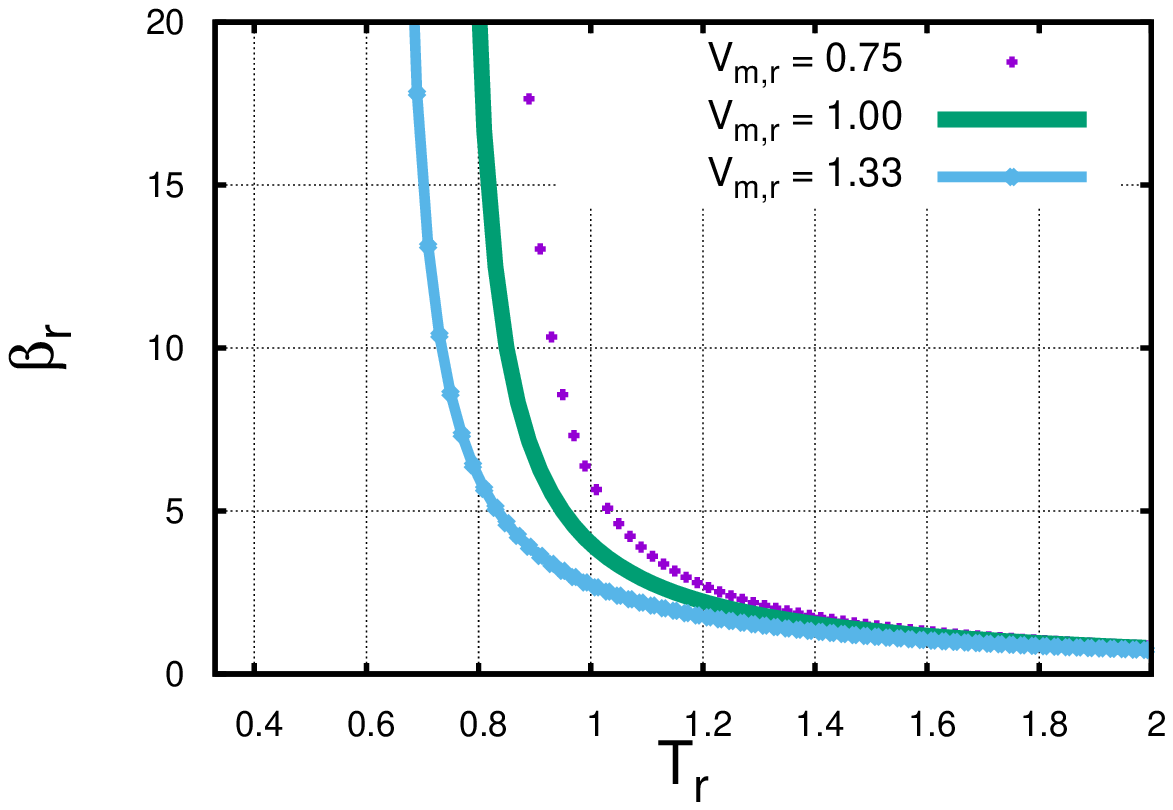}}}
\subfigure[$\beta_{r,p}$ over $\pred$ (acc. Eq. (\ref{equ:reztension}))]
{\resizebox{0.39\textwidth}{!}
{\includegraphics{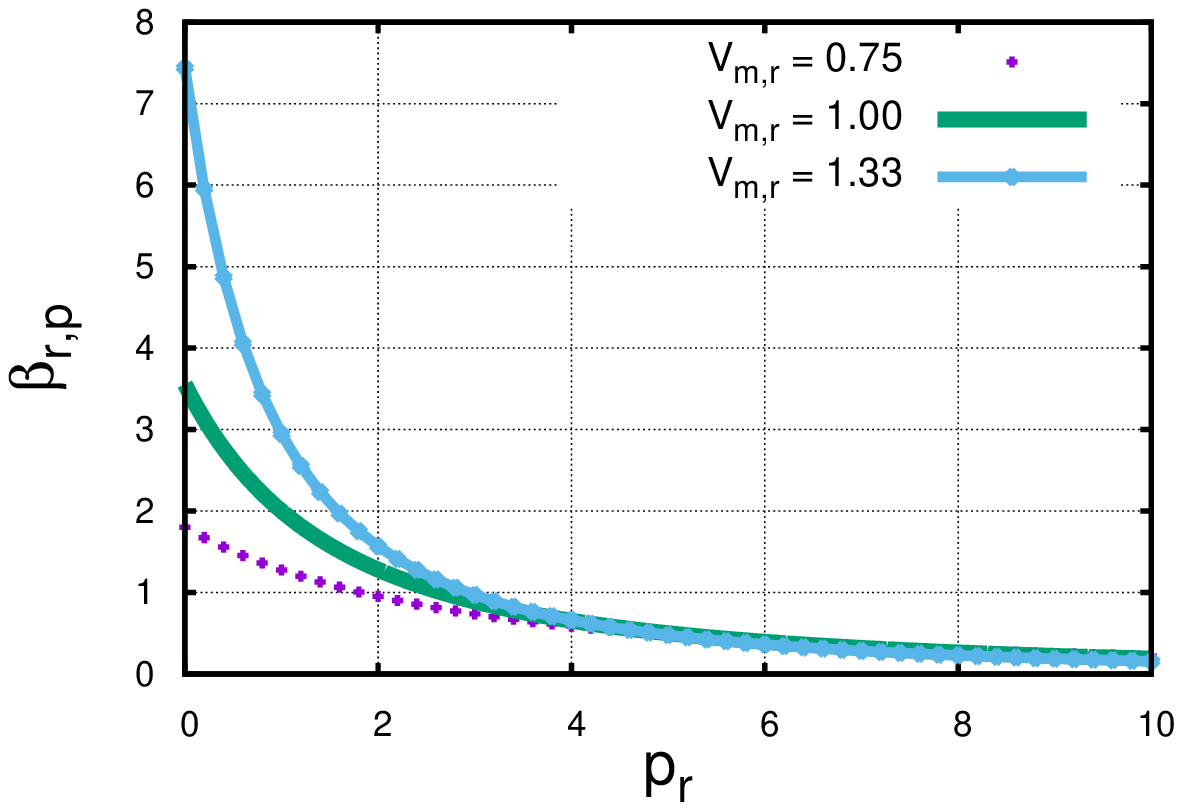}}}
\end{center}
\caption{The isochor subsets, calculated analytically by Eq.s (\ref{equ:vdw-reduced}),
(\ref{equ:tension}) and (\ref{equ:reztension})}
\label{fig:isochor}
\end{figure*}

~\nocite{sw:gnuplot-2004}

\subsubsection{The first isochor subset - Eq. (\ref{equ:dqdt-isochor-1})}

The reduced tension coefficient is:

\begin{equation}
\label{equ:tension}
\beta_{r} = \frac{1}{\pred} \frac{d \pred}{d \Tred} 
= \frac{1}{\pred} \frac{8}{(3 \Vred - 1)}
\end{equation}

The situation looks always similar: The $\beta_{r}$ value has no roots and no poles
due to the fact that the denominator is always greater than zero. That means that the 
corresponding isochor dynamics simulations runs stable independent from
the choice of $\Vred$ (see Eq. (\ref{equ:tension}) and Fig.~\ref{fig:isochor}(c)).

The analytical integration results in:

\begin{eqnarray}
\label{equ:pTfirstintegral}
\prn & = & \predi + \int_{\Tredi}^{\Tredn} \frac{8}{(3 \Vred - 1)} d \Tred 
\nonumber \\
\predn & = & \predi + \frac{8}{(3 \Vred - 1)} (\Tredn - \Tredi)
\end{eqnarray}

It can be derived according Eq.s (\ref{equ:consty}) and (\ref{equ:ykdyt})

\begin{equation}
\label{equ:simanneal}
\Tredn = \Tredi + k_{dT} \Delta t_M
\end{equation}

what is equivalent to the linear SA approach, which are implemented within
a lot of software packages like GROMACS~\cite{sw:gmx465-2013}, LAMMPS~\cite{sw:lammps-2014},
GAMESS~\cite{sw:gamess-1993}, GAUSSIAN~\cite{sw:gaussian-2009},
 CPMD~\cite{JK-CPMD-inbook-2006}, ESPReSo~\cite{sw:espresso-2013}
or Monte Carlo methods like BFM~\cite{th:binder-1979,sw:bfm-1993} and many
others.

With Eq. (\ref{equ:xydifference}) and (\ref{equ:numericxt}) follows:

\begin{equation}
\label{equ:pTnumerical}
\predn \approx \predi + \Bigg({\frac{8}{(3 \Vred - 1)}}\Bigg)_{\Vred} 
k_{dT} \Delta t_M 
\end{equation}

What can be seen here is the overall stability applying the simulated
annealing approach according Eq. (\ref{equ:simanneal}) on isochor
conditions because the factor of the second term is proportional
to the tension coefficient $\beta_{r}$.

\subsubsection{The second isochor subset - Eq. (\ref{equ:dqdt-isochor-2})}.

The pendant to the reduced tension coefficient can be derived to:

\begin{equation}
\label{equ:reztension}
\beta_{r,p} = \frac{1}{\Tred} \frac{d \Tred}{d \pred} 
= \frac{(3 \Vred - 1)}{8 \Tred}
\end{equation}

, which has positive values too in every case.

Taking the correlation of Eq. (\ref{equ:derivative-inverse}) into account
the following relation between $\beta_{r}$ and
$\beta_{r,p}$ can be formulated:

\begin{equation}
\Tred \beta_{r,p} \pred \beta_{r} = 1.
\end{equation}

The considerations can be continued with the analytic integration (see
Fig.~\ref{fig:isochor}(d)):

\begin{eqnarray}
\label{equ:Tpfirstintegral}
\Tredn & = & \Tredi + \int_{\predi}^{\predn} \frac{(3 \Vred - 1)}{8} d \pred 
\nonumber \\
\Tredn & = & \Tredi + \frac{(3 \Vred - 1)}{8} (\predn - \predi)
\end{eqnarray}

Defining a constant $k_{dp}$ according Eq. (\ref{equ:consty}) and
(\ref{equ:ykdyt}), the following expression

\begin{equation}
\label{equ:simcompress}
\predn = \predi + k_{dp} \Delta t_M 
\end{equation}

leads directly to the linear SC approach (see Zhao~\cite{JZ-simulated-compression-article-2013}).
The corresponding numerical integration method looks like:

\begin{equation}
\label{equ:Tpnumerical}
\Tredn \approx \Tredi + \Bigg(\frac{(3 \Vred - 1)}{8}\Bigg)_{\Vred} 
k_{dp} \Delta t_M 
\end{equation}

It can be stated that the analytical (Eq. (\ref{equ:Tpfirstintegral}))
and the numerical integral solution (Eq. (\ref{equ:Tpnumerical})) 
show both a stable behavior over the whole
range of reduced temperature $\Tred$, reduced pressure $\pred$ and time $t_M$.
The reduced pressure and the reduced temperature are correlating directly
like it could be expected and how it was simulated else were (see 
Neimark~\cite{ev:neimark-2005}, Medeiros~\cite{MM-GEMC-water-article-1997} and
Baidakov~\cite{VB-md-simulation-isotherm-crystallization-article-2010},
cf. with Fig.~\ref{fig:isochor}(a) and (b)). 

\bigskip

The case of isochor conditions results
in two different approaches, which are well known from the literature 
(linear SA~\cite{md:kirkpatrick-1983} and linear 
SC~\cite{JZ-simulated-compression-article-2013}).

\subsection{The isobar case}
\label{subsec:isobar}

The set of Eq. (\ref{equ:dqdt-T}), (\ref{equ:dqdt-p}) and
(\ref{equ:dqdt-V}) including the condition $\frac{d \pred}{d t_M} = 0$
gives two isobar subsets:

\begin{eqnarray}
\label{equ:dqdt-isobar-dV}
\frac{d \Vred}{d t_M} & = & - \Bigg(\frac{\frac{\delta F}{\delta \Tred}}
{\frac{\delta F}{\delta \Vred}}\Bigg)_{\pred} \frac{d \Tred}{d t_M}\\
 & = & - \frac{\dFdTlong }{\dFdVmlong } \frac{d \Tred}{d t_M}\nonumber
\end{eqnarray}

\begin{eqnarray}
\label{equ:dqdt-isobar-dT}
\frac{d \Tred}{d t_M} & = & - \Bigg(\frac{\frac{\delta F}{\delta \Vred}}
{\frac{\delta F}{\delta \Tred}}\Bigg)_{\pred} \frac{d \Vred}{d t_M}\\
 & = & - \frac{\dFdVmlong}{\dFdTlong } \frac{d \Vred}{d t_M}\nonumber
\end{eqnarray}

The constant reduced pressure $\pred$ can be chosen as a subcritical ($\pred = 0.75$),
a critical ($\pred = 1.00$) or a supercritical ($\pred = 1.33$) parameter.

\begin{figure*}[ht]
\begin{center}
\subfigure[$\Vred$ over $\Tred$ (acc. reciprocal Eq. (\ref{equ:TVmnumerical}))]
{\resizebox{0.39\textwidth}{!}
{\includegraphics{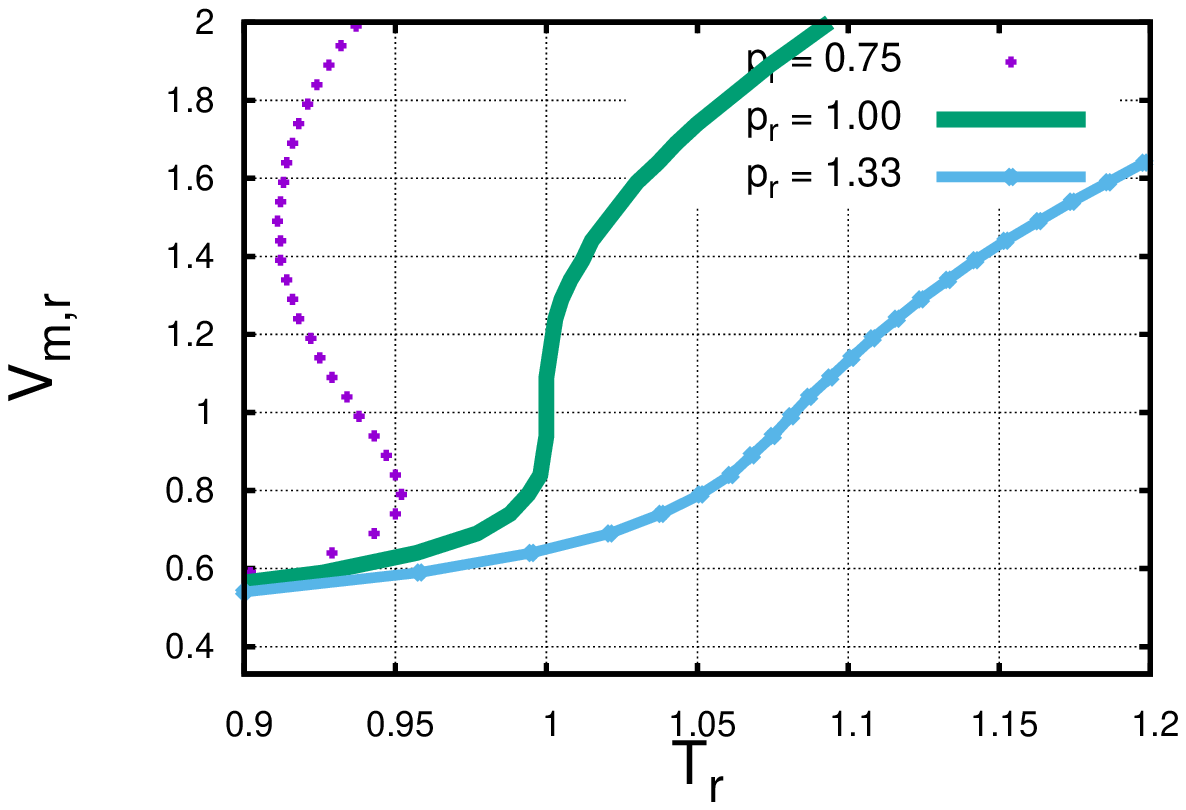}}}
\subfigure[$\Tred$ over $\Vred$ (acc. Eq. (\ref{equ:TVmnumerical}))]
{\resizebox{0.39\textwidth}{!}
{\includegraphics{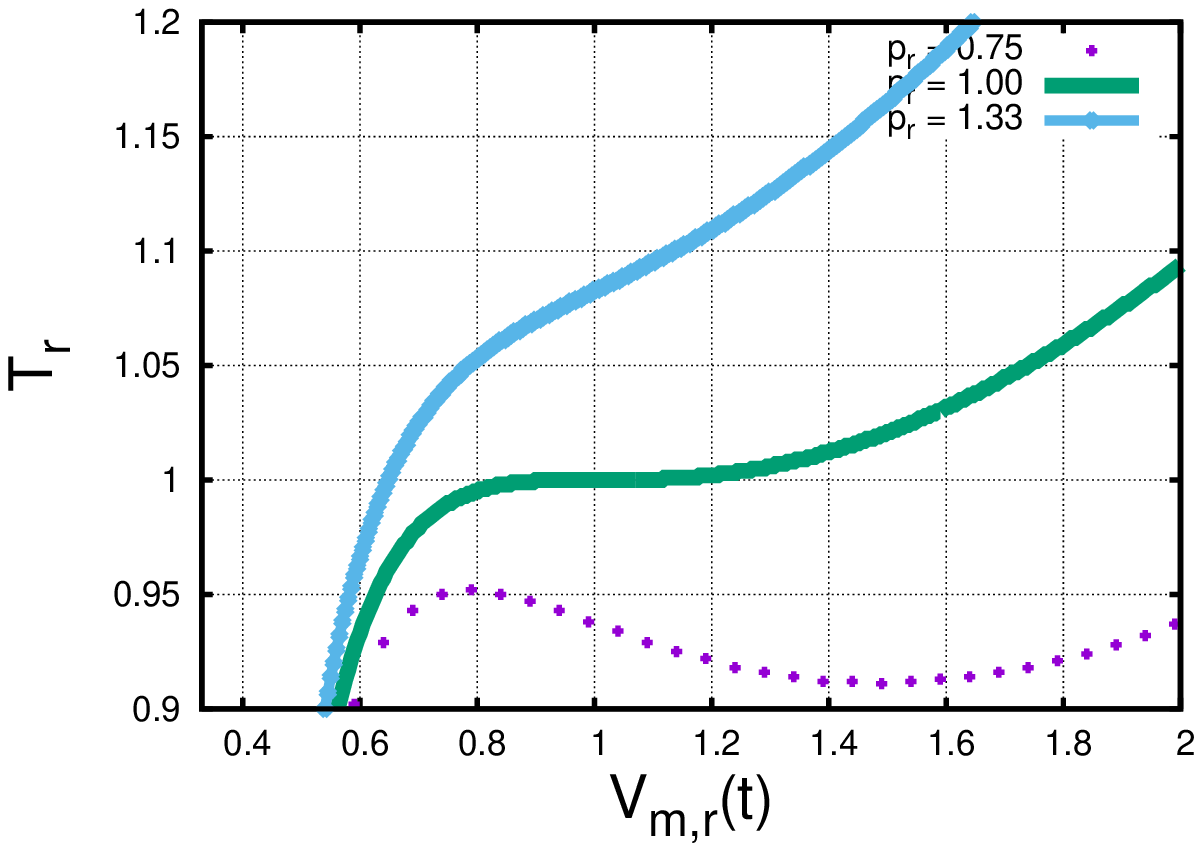}}}
\subfigure[$\alpha_r$ over $\Tred$ (acc. Eq. (\ref{equ:alpha}))]
{\resizebox{0.39\textwidth}{!}
{\includegraphics{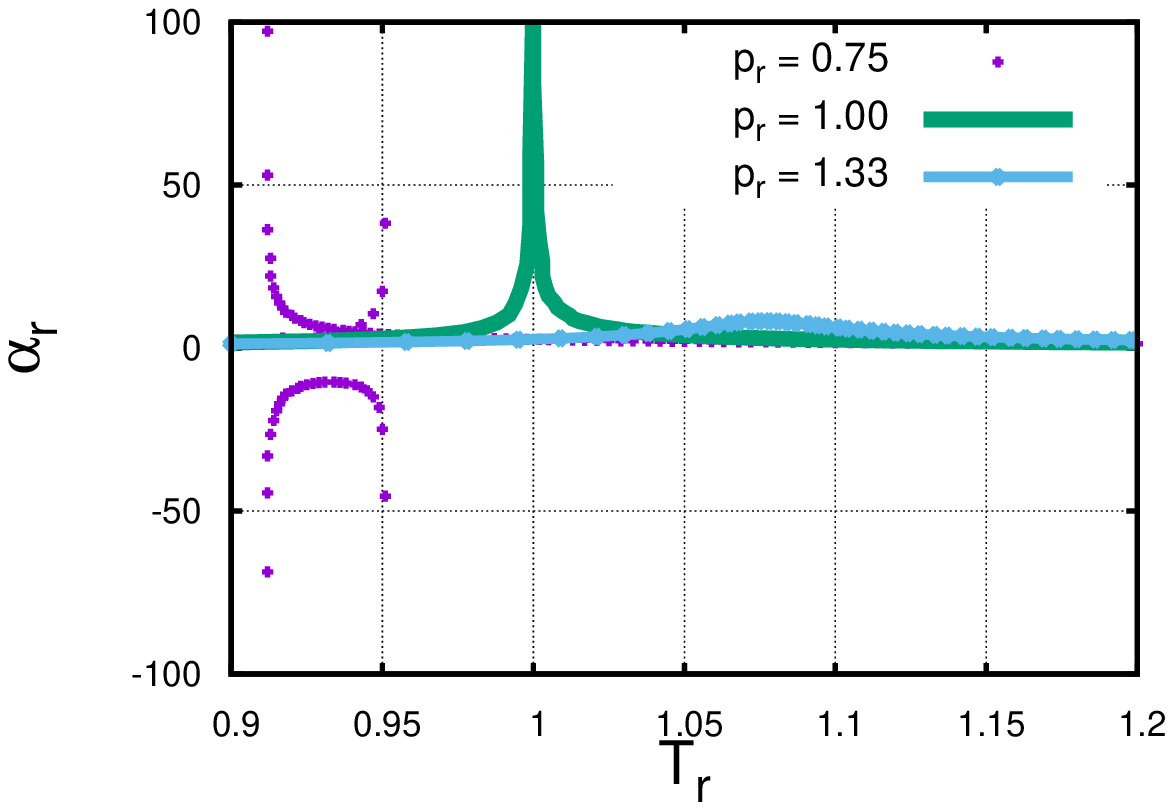}}}
\subfigure[$\alpha_{r,p}$ over $\Vred$ (acc. Eq. (\ref{equ:rezexpansion}))]
{\resizebox{0.39\textwidth}{!}
{\includegraphics{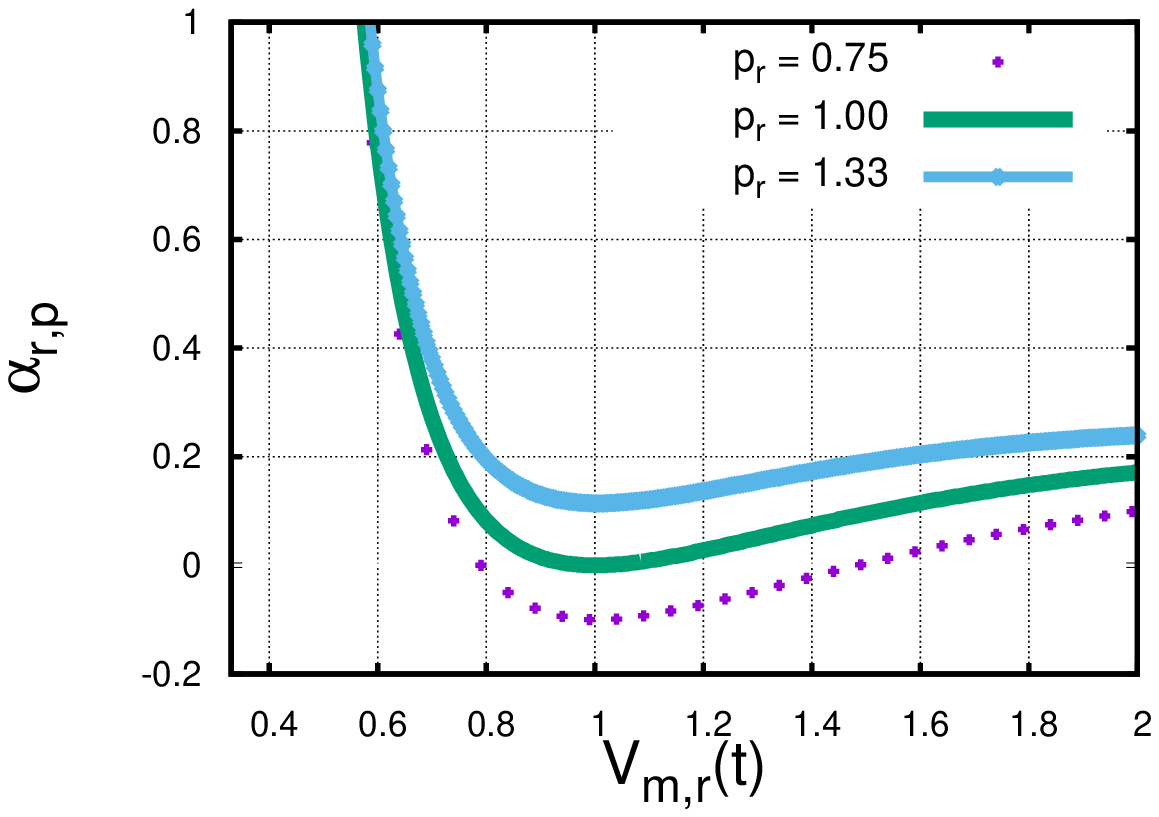}}}
\end{center}
\caption{The isobar subsets, calculated numerically by Eq.s (\ref{equ:TVmnumerical}),
(\ref{equ:rezexpansion}), (\ref{equ:alpha}), applying $k_{dV,T} = 10^{-6}$
(for reducing the error term $O(\Delta t_M)$ according Eq. (\ref{equ:xydifference}))}
\label{fig:isobar}
\end{figure*}

\subsubsection{The first isobar subset - Eq. (\ref{equ:dqdt-isobar-dV})}

The corresponding reduced susceptibility coefficient of expansion is defined by:

\begin{equation}
\label{equ:expansion}
\alpha_r = \frac{1}{\Vred} \frac{d \Vred}{d \Tred} 
= \frac{8 \Vred}{\dFdVmlong}
\end{equation}

The denominator of Eq. (\ref{equ:expansion}) has two roots in the considered
subcritical case, i.e, the fracture and therefore the
expansion coefficient has two pole values. In other words, the $\Vred$ depends on
$\Tred$ in an ambiguous manner (see Fig.~\ref{fig:isobar}(a)). However,
the reduced expansion coefficient $\alpha_r$ shows the following behavior on
the subcritical condition (see Fig.~\ref{fig:isobar}(c)): 
\begin{enumerate}
\item increasing from positive up to infinity with increasing $\Tred$, 
\item jumping to negative infinity, 
\item with decreasing $\Tred$ growing up from negative infinity to a negative
maximum and back to negative infinity,
\item jumping to positive infinity, 
\item with increasing $\Tred$ the $\alpha_r$ decreases monotonically to a positive
minimum.
\end{enumerate}

The denominator of Eq. (\ref{equ:expansion}) has one root in the considered
critical case, i.e, there is one pole value. Fig.~\ref{fig:isobar}(c)
shows, that $\alpha_r$ goes to positive infinity at the critical point.

Applying $\pred = 1.33$, the $\Vred$ shows monotonic increasing, and the 
$\alpha_r$ increases now to a limited maximum with a further decreasing.

The case of interest here is the subcritical one because of one important
question: What does an infinite expansion coefficient mean? The
mathematical point of view to Eq. (\ref{equ:expansion}) is helpful, here,
which offers two different opportunities of explanation. 

The first explanation is that $d ln \Vred$ runs up to $\infty$ and 
$d \Tred$ has a constant value $\not= 0$. Of course,  $d \Tred \not= 0$
means the linear SA method (Eq. (\ref{equ:simanneal}). According the 
shown behavior (see Fig.~\ref{fig:isobar}(c)) this results in infinite
volume fluctuations, i.e., $d ln \Vred \rightarrow \infty$, or: the obtainings
are in agreement with the simulation results, given by Walser~\cite{RW-md-simulation-water-article-2000},
Dou~\cite{md:dou-2001,md:dou-2-2001} and Zahn~\cite{md:zahn-2004}. 
But, this stands in contradiction with the experimental findings.

The other opportunity to explain the infinite expansion coefficient is that
$d ln \Vred$ has a constant limited value $\not= 0$ and
$d \Tred = 0$. This gives infinity too, of course.

A practical point of view, i.e., that such transitions are observable in
nature, e.g., the above called water cooking, prefer clearly the second
exposition. Therefore, the behavior of $\alpha_r$ can be interpreted as follows:
it is a finite limit of the speed of volume expansion coupled by a 
temperature change of zero. Unfortunately, such a point of view is unknown
from the literature of simulating phase transition
processes on the given isobar condition.

\subsubsection{The second isobar subset - Eq. (\ref{equ:dqdt-isobar-dT})}

Fortunately, Eq. (\ref{equ:dqdt-isobar-dT}) as the inverse
function to Eq. (\ref{equ:dqdt-isobar-dV}) opens up a
quite different procedure, where the reduced temperature $\Tred$ depends
on the reduced volume $\Vred$.

The corresponding susceptibility coefficient of Eq. (\ref{equ:dqdt-isobar-dT})
is the pendant of the reduced expansion coefficient, called $\alpha_{r,p}$:

\begin{equation}
\label{equ:rezexpansion}
\alpha_{r,p} = \frac{1}{\Tred} \frac{d \Tred}{d \Vred} = \frac{1}{\Tred} 
\frac{\dFdVmlong}{8 \Vred^2}
\end{equation}

The behavior of $\alpha_{r,p}$ is similar to that of an asymmetric quadratic
parable (see Fig.~\ref{fig:isobar}(d)), i.e., it comes from finite positive
values down to a minimum and increases again to further positive values with
always increasing $\Vred$. $\alpha_{r,p}$ has two roots in the subcritical, 
one root in the critical and no root in the supercritical case and no poles
at all. This corresponds with the $\Tred(\Vred, \pred = const.)$
function as an unambiguous one which behaves like a cubic parable (see
Fig.~\ref{fig:isobar}(b)), i.e., it has two (subcritical), one (critical)
or zero (supercritical) extreme values.

For completeness, the relation between $\alpha_r$
and $\alpha_{r,p}$ is given by (Eq. (\ref{equ:derivative-inverse})):

\begin{equation}
\label{equ:alpha}
\Tred \alpha_{r,p} \Vred \alpha_r = 1.
\end{equation}

Now, it is easy to integrate analytically and numerically.

The analytic integration gives (Mathematica~\cite{ma:mathematica-1997}):

\begin{equation}
\label{equ:TVmfirstintegral}
d \Tredn = \int_{\Vredi}^{\Vredn} \Bigg(
\frac{9 \pred}{8} - \frac{(\pred + 8 \Tred)}{4 \Vred} + \frac{9}{8 \Vred^2}
\Bigg)_{\pred} d \Vred
\end{equation}

\begin{eqnarray}
\label{equ:TVmfirstintegralsolution}
\Tredn & = & \Tredi \nonumber \\
& + & \frac{9 \pred}{8} (\Vredn - \Vredi) \nonumber \\
& + & \frac{(\pred + 8 \Tredi)}{4} ln \Bigg(\frac{\Vredi}{\Vredn} \Bigg) \nonumber \\
& + & \frac{9}{8} \Bigg(\frac{1}{\Vredi} - \frac{1}{\Vredn}\Bigg) \nonumber \\
\end{eqnarray}

The curvature of the temperature can be discussed now. In 
Eq. (\ref{equ:TVmfirstintegralsolution}), the second and 
the fourth term deliver always positive parts, the third term results always to be negative. 
In summary, the behavior of the temperature development is always stable with a continuous
or step by step volume change in positive or negative direction.

The numerical integration may be performed with

\begin{equation}
\label{equ:dVdtconstant}
\frac{d \Vred}{d t_M} = const. = k_{dV,T}
\end{equation}

what gives

\begin{equation}
\label{equ:simexpans}
\Vredn = \Vredi + k_{dV,T} \Delta t_M
\end{equation}

which might be called a ''Simulated Expansion'' (SE) approach. Index $T$ stands
for temperature control.

The numerical integration according to Eq. (\ref{equ:dVdtconstant}) and
(\ref{equ:simexpans}) looks like:

\begin{equation}
\label{equ:TVmnumerical}
\Tredn \approx \Tredi  + \Bigg(\frac{9 \pred}{8} - \frac{(\pred + 8 \Tredi)}{4 \Vredi} 
+ \frac{9}{8 \Vredi^2}\Bigg)_{\pred} 
k_{dV,T} \Delta t_M 
\end{equation}

, what shows an interesting behavior. Using a short description of

\begin{equation}
\label{equ:dSA}
\frac{d \Tred}{d t_M} \sim \Tred
\end{equation}

it is clear what this approach means. The control of the time derivative of the
temperature follows the temperature itself. Let us call it a 
''differential Simulated Annealing'' (dSA) method, when applying a constant 
volume change (''Simulated Expansion'' - SE). Eq. (\ref{equ:TVmnumerical}) 
is the Euler solution of an inhomogeneous differential equation of first
order by the temperature, which allows the passing of the phase transition
point of evaporation/condensation on isobar condition in the forward and
backward direction,  i.e., a cyclic (hysteresis) 
simulation is also possible.

Additionally, this equation confirms the explanation that the singularity
of the reduced expansion coefficient $\alpha_r$ matches indeed the behavior of 
$d ln \Vred = const. \not= 0$ and $d \Tred = 0$. I.e., the natural process of
the evaporation/condensation of a subcritical liquid runs with a limited speed of
volume expansion including a stop of the temperature increasing at the 
phase transition point.

\bigskip

So, Eq. (\ref{equ:TVmnumerical}) opens up the opportunity to control the macroscopic
temperature $\Tr$ in the analytic case of the VdW-EoS. Because the 
important issues of the excluded volume and the interactions between the particles
are also included in many other analytical or numerical simulation techniques, 
it can be assumed that the corresponding equations of state behave quite similar
to the Van der Waals equation of state. Therefore, Eq. (\ref{equ:dSA})
can be supposed to be valid in general for controlling $\Tr$ in Eq. (\ref{equ:berendsen34}).

\subsection{The isotherm case}
\label{subsec:isotherm}

The set of Eq.s (\ref{equ:dqdt-T}), (\ref{equ:dqdt-p})
and (\ref{equ:dqdt-V}) including the condition $\frac{d \Tred}{d t_M} = 0$
gives two isotherm subsets:

\begin{eqnarray}
\label{equ:dqdt-isotherm-dV}
\frac{d \Vred}{d t_M} & = & - \Bigg(\frac{\frac{\delta F}{\delta \pred}}
{\frac{\delta F}{\delta \Vred}}\Bigg)_{\Tred} \frac{d \pred}{d t_M}\\
 & = & - \frac{\dFdplong }{\dFdVmlong } \frac{d \pred}{d t_M} \nonumber
\end{eqnarray}

\begin{eqnarray}
\label{equ:dqdt-isotherm-dp}
\frac{d \pred}{d t_M} & = & - \Bigg(\frac{\frac{\delta F}{\delta \Vred}}
{\frac{\delta F}{\delta \pred}}\Bigg)_{\Tred} \frac{d \Vred}{d t_M}\\
 & = & - \frac{\dFdVmlong}{\dFdplong } \frac{d \Vred}{d t_M}\nonumber
\end{eqnarray}

The constant reduced temperature $\Tred$ as a simulation parameter can be set
to a subcritical value of $\Tred = 0.85$, a critical value of $\Tred = 1.00$ and to a
supercritical value of $\Tred = 1.20$.

\begin{figure*}[ht]
\begin{center}
\subfigure[$\Vred$ over $\pred$ (acc. reciprocal Eq. (\ref{equ:pVmnumerical}))]
{\resizebox{0.39\textwidth}{!}
{\includegraphics{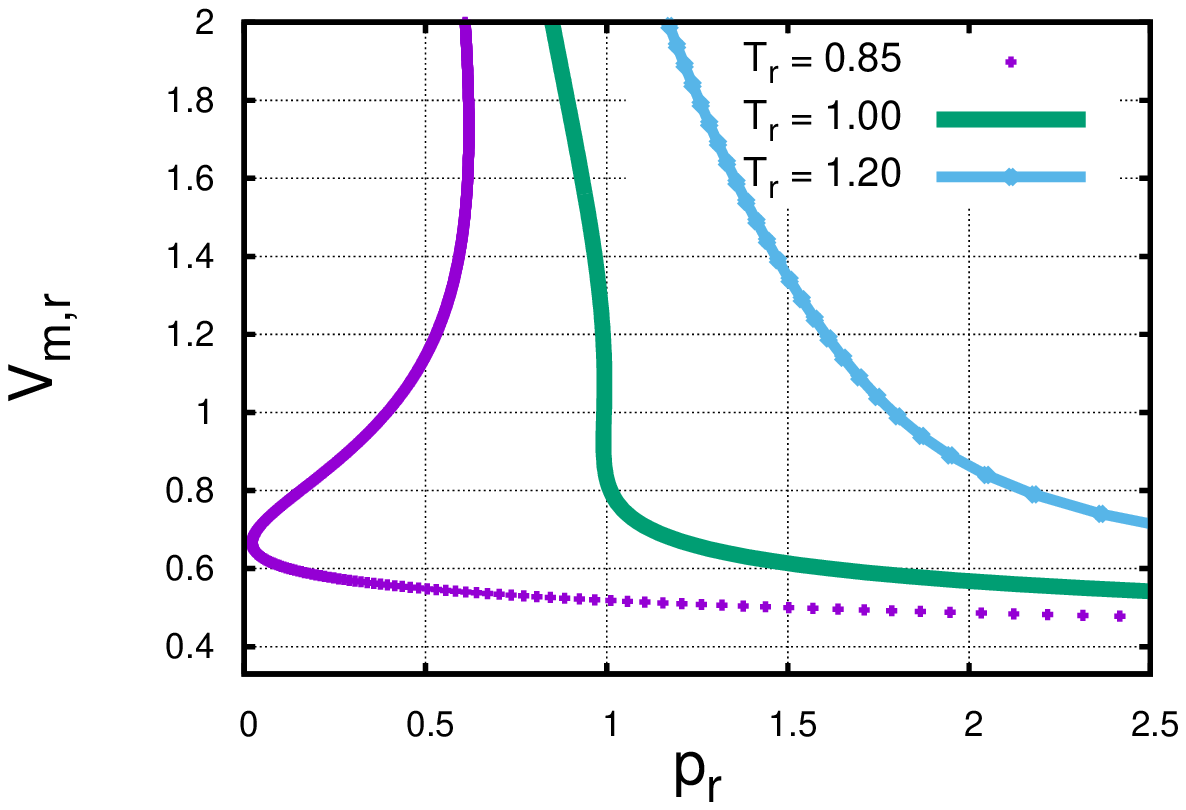}}}
\subfigure[$\pred$ over $\Vred$ (acc. Eq. (\ref{equ:pVmnumerical}))]
{\resizebox{0.39\textwidth}{!}
{\includegraphics{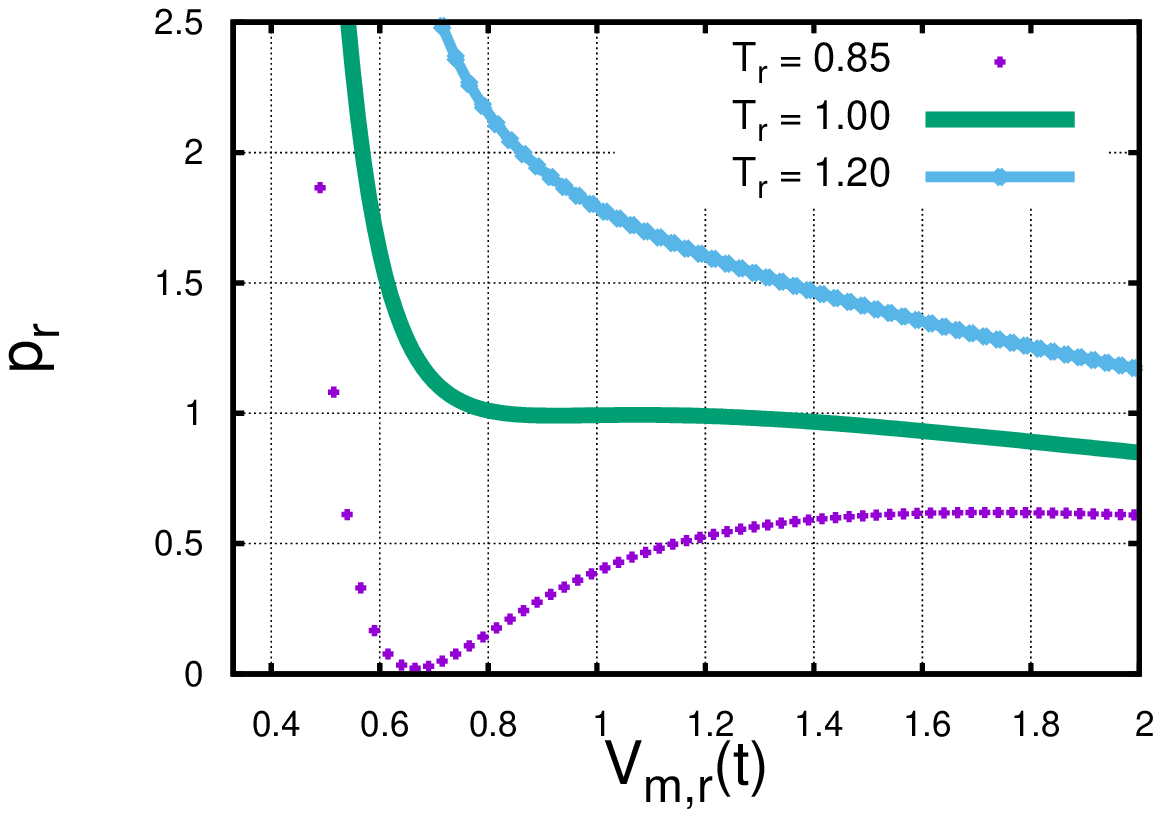}}}
\subfigure[$\kappa_r$ over $\pred$ (acc. Eq. (\ref{equ:compressibility}))]
{\resizebox{0.39\textwidth}{!}
{\includegraphics{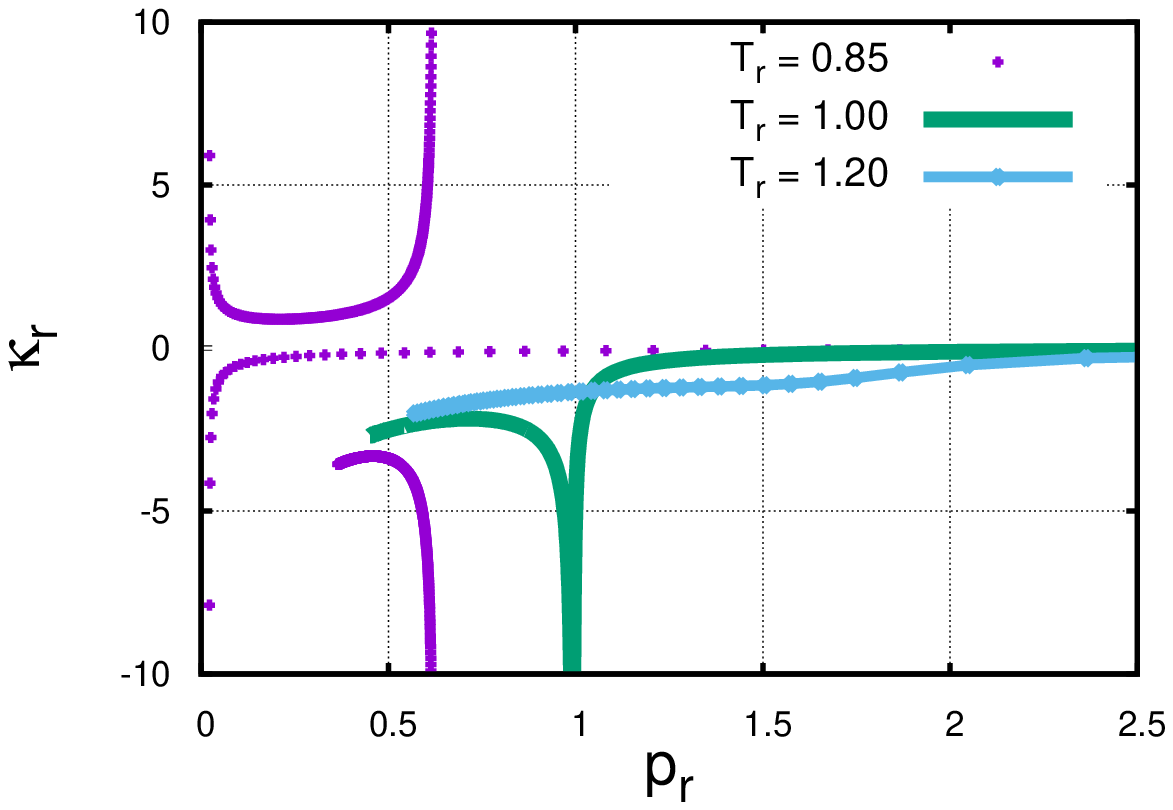}}} 
\subfigure[$\kappa_{r,p}$ over $\Vred$ (acc. Eq. (\ref{equ:rezcompressibility}))]
{\resizebox{0.39\textwidth}{!}
{\includegraphics{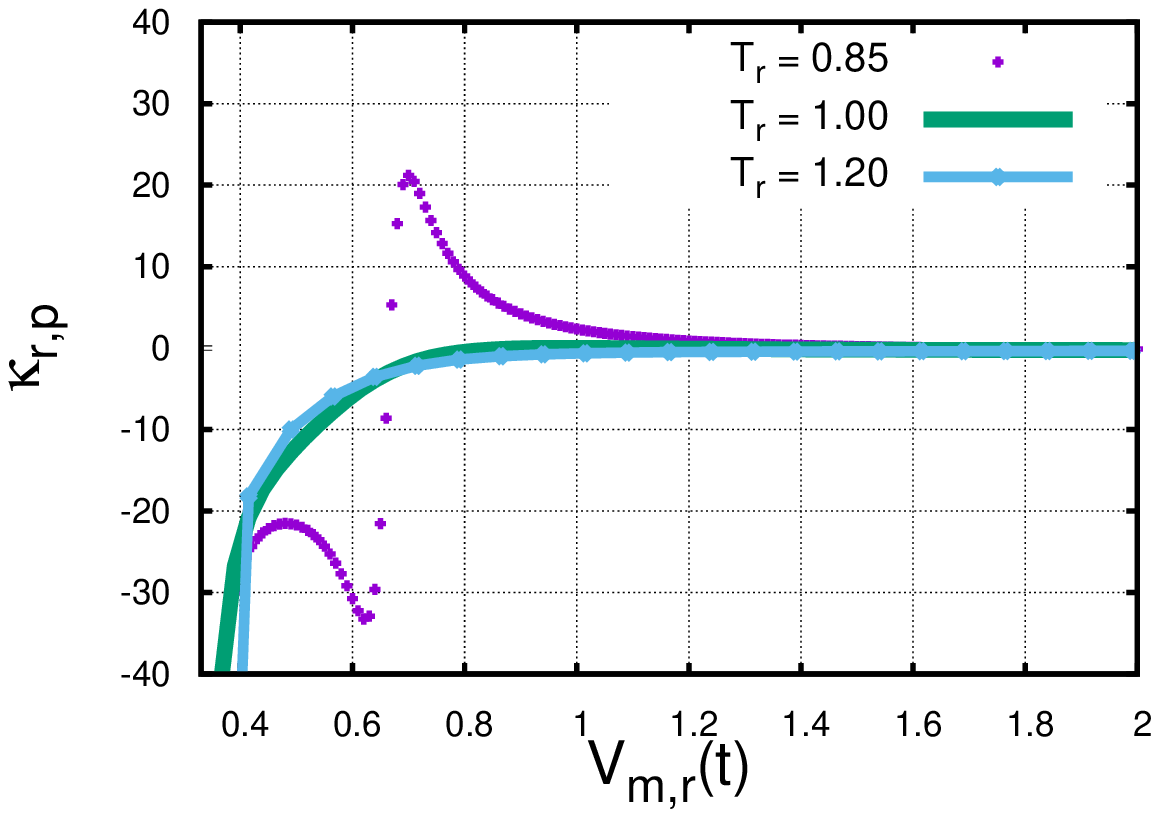}}}
\end{center}
\caption{The isotherm subsets, calculated numerically by Eq.s (\ref{equ:pVmnumerical}),
(\ref{equ:rezcompressibility}), (\ref{equ:compressibility}), applying
$k_{dV,p} = 10^{-6}$ (for reducing the error term $O(\Delta t_M)$
according Eq. (\ref{equ:xydifference}))}
\label{fig:isotherm} 
\end{figure*}

\subsubsection{The first isotherm subset - Eq. (\ref{equ:dqdt-isotherm-dV})}

The susceptibility coefficient of this approach is the reduced compressibility
which is defined as:

\begin{eqnarray}
\kappa_r & = & - \frac{1}{\Vred} \frac{d \Vred}{d \pred} \nonumber \\
& = & - \frac{1}{\Vred} \frac{\dFdplong}{\dFdVmlong}
\end{eqnarray}

The behavior of this fracture expression depends from the denominator
analogously like the discussion in the previous subsection.

The compressibility reaches values down to negative infinity in the
subcritical case with increasing $\pred$ (see 
Fig.~\ref{fig:isotherm} (c)). It jumps from there to positive infinity
and has furthermore positive values including a minimum with decreasing
$\pred$. Then, it jumps back to negative infinity with now again increasing
$\pred$ with further monotonic increasing. This corresponds with the function 
$\Vred(\pred, \Tred = const.)$ which is an ambiguous one (see
Fig.\ref{fig:isotherm}(a)).

The conclusions are quite similar like in the isobar case:
If a si\-mu\-la\-ted comp\-res\-sing me\-thod like shown in 
Eq. (\ref{equ:simcompress}) is applied, i.e., the pressure is increasing
linearly with the time, then the volume changes go to infinite high 
fluctuations (e.g.
Luo~\cite{CL-coarse-graned-model-polymer-crystallization-article-2009}).

The critical case shows one infinite negative maximum of $\kappa_r$. The supercritical
case is characterized by a continuously increasing $\kappa_r$ discontinued
by a limited minimum.

\subsubsection{The second isotherm subset - Eq. (\ref{equ:dqdt-isotherm-dp})}

Again, fortunately, Eq. (\ref{equ:dqdt-isotherm-dp}) as the inverse function
to Eq. (\ref{equ:dqdt-isotherm-dV}) serves as a second opportunity of
performing isotherm simulations.

The corresponding susceptibility coefficient $\kappa_{r,p}$ is called:

\begin{eqnarray}
\label{equ:rezcompressibility}
\kappa_{r,p} & = & - \frac{1}{\pred} \frac{d \pred}{d \Vred} \nonumber \\
& = & - \frac{1}{\pred} 
\frac{\dFdVmlong}{\dFdplong}
\end{eqnarray}

The behavior of this coefficient is similar to that of $\alpha_{r,p}$
with negative sign. It has roots but no poles (see Fig.~\ref{fig:isotherm}(d))
independent from the parameter choice of $\Tred$.

The relation between $\kappa_r$ and $\kappa_{r,p}$ according
Eq. (\ref{equ:derivative-inverse}) is:

\begin{equation}
\label{equ:compressibility}
\pred \kappa_{r,p} \Vred \kappa_r = 1.
\end{equation}

The analytic solution of Eq. (\ref{equ:dqdt-isotherm-dp}) is given
to (Mathematica~\cite{ma:mathematica-1997}):

\begin{equation}
\label{equ:pVmfirstintegral}
d \predn = - \int_{\Vredi}^{\Vredn} \Bigg(
\frac{\dFdVmlong}{\dFdplong } \Bigg)_{\Tred} d \Vred
\end{equation}

\begin{eqnarray}
\label{equ:pVmfirstintegralsolution}
\predn & = & \predi \nonumber \\
& - & \Bigg(9 \bigg(\frac{1}{\Vredn} - \frac{1}{\Vredi}\bigg) \nonumber \\
& + & \bigg(2 \predi + 16 \Tred - 27 \bigg) ln \bigg(\frac{\Vredn}{\Vredi}\bigg) \nonumber \\
\nonumber \\
& + & \bigg(27 + \predi - 16 \Tred \bigg) ln \bigg(\frac{(3 \Vredn - 1)}{(3 \Vredi - 1)}\bigg) \Bigg)
\nonumber \\
\end{eqnarray}

This solution (Eq. (\ref{equ:pVmfirstintegralsolution})) includes parts 
which deliver positive or negative contributions to the reduced pressure $\pred$
without any point of indefiniteness. 
An increasing volume, continuously or step by step with 
$k_{dV,p}$ (see below) applied, decreases the pressure (with the exception of
passing a phase transition point) what is in agreement with the general
expectations of such a kind of equations of state.

The application of the simulated expansion method needs a new constant
definition for the numerical integration

\begin{equation}
\label{equ:dVdtconstantp}
\frac{d \Vred}{d t_M} = const. = k_{dV,p}
\end{equation}

what gives with index $p$ for the pressure control:

\begin{equation}
\label{equ:simexpansp}
\Vredn = \Vredi + k_{dV,p} \Delta t_M
\end{equation}

The numerical solution is now:

\begin{eqnarray}
\label{equ:pVmnumerical}
\predn & \approx & \predi  \nonumber \\
& - & \Bigg(\frac{(9 \predi \Vred^2 - 2 (\predi + 8 \Tred) \Vred + 9)}
{\dFdplong}\Bigg)_{\Tred} 
k_{dV,p} \Delta t_M \nonumber \\ 
\end{eqnarray}

I.e., the reduced macroscopic reference pressure $\pred$ is controlled indirectly
by a constant volume change, similar to the isobar case. 

The expression

\begin{equation}
\label{equ:dSC}
\frac{d \pred}{d t_M} \sim \pred
\end{equation}

shows the meaning of this modified simulated expansion: it is a method which
might be called ''differential Simulated Compressing'' (dSC) which is unknown
until now.

\bigskip

With the same reasoning given in the subsection above Eq. (\ref{equ:dSC}) can 
be supposed to be valid in general and implementable into the pressure bath coupling method 
in order to control $\pr$ in Eq. (\ref{equ:my}). This opens up the opportunity to simulate
phase transition processes like melting/crystallization on isotherm conditions.

\subsection{General cases of combination}
\label{subsec:general}

The two expressions Eq.s (\ref{equ:dqdt-T}) and (\ref{equ:dqdt-p}) allow 
simulations by combining different methods.

Eq. (\ref{equ:dqdt-T}) defines the temperature control by the combination 
of a differential temperature changing with the $k_{dV,T}$ constant (dSA)
and a linear pressure changing with the $k_{dp}$ constant (SC). This
temperature control can be used for providing $\Tr$ 
in Eq. (\ref{equ:berendsen34}). 

Eq. (\ref{equ:dqdt-p}) defines the pressure control defining the linear temperature
$k_{dT}$ constant (SA) and the differential pressure changing with the $k_{dV,p}$
constant (dSC). It is useable to control $\pr$ in Eq. (\ref{equ:my}).

The combination of Eq.s (\ref{equ:dqdt-T}) and (\ref{equ:dqdt-p}) including
positive or negative values of $k_{dV,T}$ and $k_{dV,p}$ enables an heat
and pressure bath coupling in a general cyclic manner, i.e., performing simulations
of hysteresis processes like the famous Carnot process are achievable.

Eq. (\ref{equ:dqdt-V}) allows the combination of $k_{dT}$ and $k_{dp}$ without
any limits.

\subsection{The models and the nature of the liquid-gas phase transition}

The Van-der-Waals equation of state deals with the real gas model by the 
introduction of the excluded volume and the particle interaction. This was an 
expansion of the ideal gas equation of state in order to describe a two
phase system (liquid and gaseous) for the first time. This model equation included a phase
transition. It included also singularities of the corresponding susceptibility
coefficients and negative values of them. 

In summary it can be stated that this model describes important aspects
of the nature of a phase transition but not it's nature
itself (see Smolin~\cite{LS-universum-der-zeit-book-2014}).

Therefore, the view to the natural process itself is helpful.

The liquid-gas phase transition on isobar conditions can be observed easily by
the daily experiences. The liquid is heated up until the point of cooking. It
does not play any role if there is a spontaneous or a germ initialized evaporation
process starting at the point of phase transition. The only one fact here which is
important is that the temperature is no longer increasing until all the liquid is 
evaporated. I.e., all the heat is invested now into the transition enthalpy. In summary:
There is a jump of the whole enthalpy of the system without any change of the temperature. 
After all the liquid is evaporated the temperature is able to increase furthermore.

Maxwell~\cite{JM-maxwell-construction-1875} did argue with the same
experimental point of view when he presented his after him called construction. 
This is a linear line between two points of the isotherm curvature in order
to define an area of coexistence of liquid and gaseous phase, where the Gibbs
free energy does not change, i.e., the region is a stable one and the linear
Maxwell construction could be defined as to be both isotherm and isobar. Of
course, this agrees with the experimental observations.

Only one flaw is over now: This hypothesis could not be substantiated regarding it's
validity until now by theoretical or simulative modeling work, although, there were 
taken many efforts (Walser~\cite{RW-md-simulation-water-article-2000},
Dou~\cite{md:dou-2001,md:dou-2-2001} and Zahn~\cite{md:zahn-2004}).

There is a hope that the presented method of simulated expansion can help.

Some important questions result from these discrepancies between the theory and the
observations:

\begin{enumerate}
\item The current view is a thermodynamic one. I.e., the aspect that the phase transition
is also a process should be included into the considerations, in other
words spoken: kinetic aspects (Schwetlick ~\cite{ki:schwetlick-1975} or
Bittrich~\cite{ki:bittrich-1986} and many others) has to be respected, too. 
The following equation
\begin{eqnarray}
(A)_N \rightleftharpoons (A)_N^{\not=} \rightleftharpoons N \; A
\end{eqnarray}
where $(A)_N$ is the liquid phase, $(A)_N^{\not=}$ is a meta stable
transition complex (see the theory of transition states by Eyring
~\cite{HE-theorie-of-transition-state-article-1935}) and $N \; A$
is the gaseous phase of $N$ particles $A$, underlines the need of this time dependence
at the macroscopic level.
\item What is the difference between the gaseous and the liquid state of 
a substance of the corresponding particles apart from their inner structure? 
The gas state means that all particles carry all the translational degrees
of freedom which are frozen when the substance condensates (either to a 
liquid or a solid).
\item What is the nature of the evaporation/condensation enthalpy and the correlation
with that freezing of the translational degrees of freedom?
\item Do negative susceptibility coefficients be forbidden? New
experimentally given observations have been shown a lot of unusual behavior
of different substances: negative $\alpha_r$: Zirconium tungstate~\cite{AK-ZiW-misc-2015}
, polymers (Strobl~\cite{th:strobl-1996}, pp. 136), diblock copolymers 
(Jehnichen~\cite{DJ-blockcopolymers-thinfilms-inproceedings-2013}). Or: 
There are similar observations for the same unusual behavior corresponding 
to the compressibility coefficient (e.g., Nicolaou~\cite{ZN-negative-compressibility-article-2012}).
\end{enumerate}

\section{Summary and Outlook}
\label{sec:conc}

Molecular simulations allow so called quasi static state changes by means of
simulated annealing and simulated compressing where the molecular ensemble
is always equilibrated at the microscopic time scale $t_m$. In that sense a reasoning was
given in order to apply the time dependence of the Van der Waals equation of
state at the macroscopic time scale $t_M$ to perform these quasi static state
changes in a more general manner. Therefore, a clear distinction between these
different time scales must be respected.

Isochor, isobar and the isotherm conditions could be considered in detail 
applying some simple mathematical derivations and simplifications.

It could be shown that the simulated annealing and the simulated compressing
methods could be applied without any restrictions on isochor conditions.

It could be shown furthermore that the application of the simulated annealing 
approach to a subcritical model ensemble on isobar conditions did lead to a 
situation where the expansion coefficient reached a singularity point. The volume
as a dependant of the linear temperature increasing did run into infinite high fluctuations. 
Vice versa it could be shown by the application of the corresponding inverse function (the 
temperature as the dependant of the volume) 
that the passing of the phase transition point is a process of continuous or step by step
volume increasing. The temperature was able to reach a maximum point including
a further decreasing down to a minimum followed by an increasing again. This
is a much more better explanation for the appearance of singularities of $\alpha_r$
when passing a phase transition point because it agrees with the observations
of the natural process. This new procedure was called the simulated expansion
approach what could be shown to be equal to a differential simulated annealing method.

The application of the simulated compressing approach to a subcritical model
ensemble on isotherm conditions did lead to a singular behavior of the 
compressibility coefficient. The volume as a dependant of the pressure did
run into infinite high fluctuations.
The usage of the corresponding inverse function, i.e., the pressure is depending
on the volume, opened up another opportunity to pass such a point of singularity
of the susceptibility coefficient by a continuous or step by step volume increasing. The pressure
as a dependant of the volume was able to reach a minimum including a further 
increasing up to a maximum followed by an decreasing again. This method
was also called a simulated expansion, now on isotherm conditions, which was
proven to be a differential simulated compressing approach.

It could be shown too that a generalization of all these methods is possible. 
Analytic and numerical simulation techniques which are including the issues
of the exclude volume and the particle interaction correspond with equations
of state which are familiar with that of Van der Waals. Therefore, it can be
assumed a similar behavior what justifies the implementation of these
approaches into corresponding software packages for testing.

The continuous or step by step volume alteration by $k_{dV,(p|T)}$ of the simulation box
has no restriction to it's sign. This makes the simulation of circular
processes feasible.

\section{Acknowledgments}

The author would like to say thank to Lutz Peitzsch for many help-full
discussions for finding hints and the right way doing this work. 
These acknowledgments are also more than valid to a good friend and
too early gone away college, J{\"o}rg Bergmann, TU Dresden. 
The author would like to say thank also to Christina Claus,
Luxembourg, for psychological support.

\bibliographystyle{unsrt}

\end{document}